# Spintronic Bayesian Hardware Driven by Stochastic Magnetic Domain Wall Dynamics

**Authors**

Tianyi Wang[1,*,†], Bingqian Dai[1,*,†], Kin Wong[1], Yaochen Li[1], Yang Cheng[1], Qingyuan Shu[1], Haoran He[1], Puyang Huang[1] , Hanshen Huang[1] and Kang L. Wang[1,†]

**Affiliations**

*[1]Department of Electrical and Computer Engineering, University of California, Los Angeles, California 90095, United States*

*These authors contributed equally to this work.

Corresponding author E-mail: [†]tianyiwang0220@g.ucla.edu, [†]bdai@g.ucla.edu, [†]wang@ee.ucla.edu

**Abstract**

As artificial intelligence (AI) advances into diverse applications, ensuring the reliability and robustness of AI models is increasingly critical [1,2]. Conventional neural networks offer strong predictive capabilities but produce deterministic outputs without inherent uncertainty estimation, limiting their reliability in safety-critical domains such as autonomous driving and medical diagnostics. Probabilistic neural networks (PNNs), which introduce randomness into weights, activations, or network dynamics, have emerged as a powerful approach for enabling intrinsic uncertainty quantification [3–7]. However, traditional CMOS architectures are inherently designed for deterministic operation, actively suppressing intrinsic randomness from sources such as thermal fluctuations. This poses a fundamental challenge for implementing PNNs, where the need for extensive random number generation and probabilistic processing introduces significant computational overhead, limiting their practical deployment on conventional hardware platforms. To address this challenge, we introduce a Magnetic Probabilistic Computing (MPC) platform—an energy-efficient, scalable hardware accelerator that leverages intrinsic magnetic stochasticity for uncertainty-aware computing. This physics-driven strategy utilizes spintronic systems based on magnetic domain walls (DWs) [8–12] and their inherent dynamics [13–15] to establish a new paradigm of physical probabilistic computing for AI applications. The MPC platform integrates three key mechanisms: thermally induced DW stochasticity, voltage-controlled magnetic anisotropy (VCMA) [16,17], and tunneling magnetoresistance (TMR) [18,19], enabling fully electrical and tunable probabilistic functionality at the device level. As a representative demonstration, we implement a Bayesian Neural Network (BNN) inference structure and validate its functionality through CIFAR-10 classification tasks. Compared to standard 28 nm CMOS implementations, our approach achieves a seven-orders-of-magnitude improvement in the overall figure of merit, demonstrating substantial gains in area efficiency, energy consumption, and speed. These results underscore the potential of the MPC platform and open new hardware avenues toward reliable and trustworthy physical AI systems.

## Main text

### Introduction

The rapid advancement of artificial intelligence has driven its widespread adoption across diverse fields. As AI systems are increasingly deployed in applications that impact human lives, ensuring their reliability, robustness, and trustworthiness has become a paramount concern [1,2]. Conventional deep learning models—particularly fixed-weight neural networks—have achieved remarkable success. However, despite their great achievements, these models operate in a purely deterministic manner, producing point estimates without any quantification of uncertainty. This lack of uncertainty awareness poses a significant limitation, especially in safety-critical or data-scarce environments, where overconfident but incorrect predictions can have serious consequences. To address this limitation, PNNs have emerged as a compelling alternative [3–7]. PNNs incorporate uncertainty by treating certain components of the network as probability distributions rather than fixed values. This probabilistic framework enables PNNs to quantify prediction uncertainty, thereby enhancing model interpretability. In addition, PNNs exhibit improved resilience to overfitting and adversarial perturbations, further enhancing their reliability in real-world applications. However, the practical deployment of PNNs remains challenging due to their high computational complexity. Probabilistic inference typically requires repeated sampling from posterior or intermediate distributions, substantially increasing runtime and energy consumption compared to conventional neural networks. This challenge is particularly pronounced in traditional CMOS architectures, primarily due to the high demand for random number generation. Addressing this computational bottleneck is essential for unlocking the full potential of probabilistic inference in next-generation AI systems. Although various hardware-based simulated solutions have been proposed to address this challenge [20-23], very few experimental prototypes have been successfully demonstrated.

To address this bottleneck, we propose a physics-driven computing strategy that harnesses emerging spintronic systems based on magnetic solitons—such as DWs [8,9] and skyrmions [10–12]—to exploit intrinsic magnetic dynamics [13–15], thereby establishing a new paradigm of physical probabilistic computing tailored for AI applications. Specifically, we introduce a MPC hardware that integrates DWs in magnetic tunneling junctions (MTJs) [18,19] with VCMA [16,17], enabling the direct physical implementation of PNNs. The magnetic domains—reaching nanoscale dimensions as small as 1 nm [24]—offer a compelling pathway for device scaling. The exceptionally low energy cost of DW motion, requiring only 27 aJ per operation [25], underscores its potential for ultra-low-power computation. DW motion speed can reach up to 4000 m/s [26], which supports high-speed computing and memory operation. In addition, magnetic nonvolatility [27] enables seamless in-memory computing, reducing data transfer overhead and enhancing system performance for complex neural workloads. Importantly, the stochastic nature of DW dynamics—combined with VCMA-enabled tunable probability and TMR-enabled high-precision readout—makes this architecture inherently suitable for implementing probabilistic in-memory computing. This architecture acts as a stochastic unit capable of generating output signals that follow a tunable Gaussian distribution. This enables efficient sampling and uncertainty estimation, offering a substantial improvement over conventional methods. As a representative demonstration, we implement a BNN, a representative class of PNNs, and validate its functionality through CIFAR-10 classification tasks. Benchmarking results show that, compared to conventional 28 nm CMOS implementations, our approach achieves a seven-orders-of-magnitude improvement in the overall figure of merit, reflecting significant enhancements in area efficiency, energy consumption, and computational speed. This experimental demonstration underscores the potential of the MPC paradigm and opens new avenues toward the development of reliable and trustworthy physical AI systems.

## Results

### Magnetic probabilistic computing platform and physical implementation of tunable probability

BNNs extend conventional neural networks by representing weights as probability distributions rather than fixed point estimates. [4-6] In conventional neural networks, weights are simply deterministic and fixed values for each parameter, as depicted in Fig. 1a(i). BNNs maintain a similar architectural framework but model the weights as probability distributions rather than point estimates, as illustrated in Fig. 1a(ii). This probabilistic approach allows

BNNs to quantify predictive uncertainty, thereby improving model robustness and reliability. While traditional neural networks optimize fixed weights by maximizing the likelihood function, BNNs leverage Bayes' theorem to infer a posterior distribution over the weights given the observed training data.

Since the exact posterior $p(w|D)$ is typically intractable due to the high dimensionality and complexity of neural network models, Variational Inference (VI) [28-30] is commonly employed. VI approximates the true posterior with a simpler, tractable distribution $q(w;\phi)$, often chosen to be Gaussian, where $\phi$ denotes the mean and standard deviation parameters. The objective is to minimize the Kullback–Leibler (KL) [31] divergence between the approximate and true posteriors. This is equivalent to maximizing the Evidence Lower Bound (ELBO) [32]. During training, weights are sampled from the variational distribution using the reparameterization method [33], enabling backpropagation. The model computes the expected log-likelihood and KL divergence for each mini-batch, and the ELBO is optimized through gradient-based updates. This approach enables BNNs to learn both accurate predictions and meaningful uncertainty estimates. More details are discussed in Methods.

Based on the previous discussion, the implementation of a BNN requires generating Gaussian-distributed weights with tunable mean and standard deviation to represent uncertainty in model parameters. In conventional CMOS-based hardware, this process is nontrivial and introduces significant computational overhead. The standard approach involves multiple stages: generating base uniform random numbers, transforming these uniform values into Gaussian-distributed samples, and finally applying scaling and shifting operations using digital multipliers and adders to control the desired mean and variance. Each of these steps incurs latency, energy consumption, and silicon area costs. As a result, implementing these processes in CMOS leads to large area footprints and elevated power consumption, especially when sampling must be performed repeatedly and in parallel across large neural networks. This complexity presents a major bottleneck for deploying BNNs on traditional hardware, particularly in edge AI applications where resources are limited and energy efficiency is critical. To address this challenge, we design a Magnetic Probabilistic Computing (MPC) platform—a hardware accelerator for energy-efficient uncertainty-aware computing based on intrinsic magnetic stochasticity. The hardware architecture is shown in Fig. 1b. The proposed device is constructed on MTJ [18,19] stacks comprising three layers: the free layer, the tunnel barrier, and the fixed layer. Within the free layer, the magnetic DW acts as the stochastic source of Gaussian-distribution. Red and blue arrows denote regions of positive and negative out-of-plane magnetization, respectively, while the grey arrow indicates the in-plane magnetization direction, corresponding to the position of the DW. The top electrode, connected to the fixed magnetic layer, serves as the readout terminal. A voltage is applied vertically across the MTJ stack to measure the TMR signal. TMR arises from the spin-dependent tunneling of electrons between two ferromagnetic layers separated by an insulating barrier. The resistance of the MTJ depends on the relative magnetization orientation of the free and fixed layers: it is low when the layers are aligned in a parallel (P) configuration and high when they are antiparallel (AP). Intermediate configurations, in which the free layer magnetization is only partially aligned, yield TMR values between the P and AP states. The TMR is proportional to the net magnetization beneath the readout terminal, enabling analog readout of the magnetic state. The

TMR ratio is defined as:

$$TMR\ ratio = \frac{R_{AP} - R_p}{R_p}$$

where $R_{AP}$ and $R_p$ are the resistances in the AP and P states, respectively.

We next discuss how a Gaussian distribution can be realized based on DW dynamics. By precisely controlling the motion of the DW beneath the MTJ read terminal, the net magnetization in the sensing region can be finely tuned, enabling accurate modulation of the TMR signal. This controlled variability forms the basis for generating tunable Gaussian-distributed outputs. Fig. 1c(i) illustrates DWs positioned at distinct locations (labeled ①, ②, and ③) along the magnetic channel. Simultaneously, thermal excitations perturb the magnetic moment of the DW, causing it to fluctuate around its mean position. Fig. 1c(ii) displays a DW profile obtained from finite temperature micromagnetic simulations using the MuMax3 software [34]. The irregular bending of DW showcases the effect of thermal fluctuations, serving as the physical basis of the observed stochasticity. Fig. 1c(iii) provides an overview of the operating mechanism. The location of labels ①, ②, and ③ in Fig. 1c(iii) presents the correspondence between the TMR signal and the DW's mean position in Fig. 1c(i), enabling the control of Gaussian mean. Simultaneously, thermal fluctuations alter the net magnetization around the mean value within the MTJ readout terminal, leading to variations in the TMR signal (see Fig. 1c(iv)) and enabling the generation of Gaussian standard variation (Fig. 1c(v)). To sum up, by combining the tunability of DW position and fluctuation, the Gaussian-distributed random numbers are physically achieved. These principles will be experimentally verified in the next section.

## Experimental realization of the magnetic probabilistic computing device

We next discuss the experimental implementations of our MPC platform. The experimental realization of the MPC device can be subdivided into three categories: (1) Generating a DW by injecting current into the Oersted channel. (2) Shifting the DW under the MTJ pillar by spin-orbit torque (SOT). (3) Reading the TMR signal to generate a tunable Gaussian distribution. Fig. 2a presents the optical microscope image of the MPC device after fabrication. The detailed device structure and fabrication process are discussed in Extended Data Fig. 1 and Extended Data Fig. 2. The domain generation is realized by injecting a current pulse into the right Oersted channel which is marked in green in Fig. 2a. The detailed zoomed-in magneto-optical Kerr effect (MOKE) image is shown in Fig. 2b. Before the current injection, the magnetization within the domain channel is in -z direction. After the current injection, a $6\mu m$ domain is generated and the DW location is marked by a blue arrow. DW shifting is achieved via SOT. When a charge current flows along the magnetic strip, it generates a transverse spin current through the spin Hall effect, which in turn exerts a torque on the DW, inducing its motion [35,36]. A sequence of current pulses is applied into the domain channel to drive the DW from the right to the left. The DW motion process is captured by the MOKE images, which are presented in Fig. 2c. Fig. 2c(i) illustrates the DW motion at one specific time stamp. Fig. 2c(ii) image series captures the whole DW motion process. Up-domain and down-domain regions are shaded with red and blue for better visualization. See Methods and Extended Data Fig. 3 for details about DW generation and SOT-driven DW motion.

Tuning a Gaussian distribution involves adjusting its mean and standard deviation. As discussed in the previous section, the mean value depends on the DW position under the MTJ pillar. When the DW resides at different positions beneath the MTJ pillar (Fig. 2d), the TMR readout exhibits corresponding mean values. Fig. 2e presents the TMR readout as the DW traverses beneath the MTJ pillar. The DW motion is driven by a series of current pulses applied sequentially, which incrementally shift the DW position along the domain channel. During this process, the TMR signal is continuously measured in real time. As the DW moves, the local magnetization beneath the readout head changes, leading to a gradual variation in the TMR value. Since the motion is driven by timed current pulses and the TMR is recorded continuously during and between these pulses, the x-axis is plotted on a time scale to reflect the temporal evolution of the TMR signal as the DW moves past the MTJ. The TMR readouts at positions ①, ②, and ③ correspond to the respective DW locations shown in Fig. 2d. If we stop the DW at a certain location, its mean position should be centered at corresponding positions on the TMR loop. We then stop the DW at each of the three target positions shown in Figs. 2d and 2e and repeatedly collect 5,000 data points by continuously measuring the TMR signal over time. Due to thermal fluctuations, the DW exhibits random deviation around its mean position, resulting in corresponding fluctuations in the net magnetization beneath the readout head. This leads to stochastic variations in the TMR signal, i.e., the standard deviation of Gaussian distribution. As these fluctuations are thermally driven, the resulting TMR values are expected to follow a Gaussian distribution centered at each target DW location. The results are shown in Fig. 2f, where three distinct Gaussian distributions are observed, each with a different mean corresponding to the DW positions, but with identical standard deviations reflecting consistent thermal noise characteristics. The comparison between ideal Gaussian and our experimentally obtained ones are presented in Extended Data Fig. 4.

**Voltage-controlled standard deviation of Gaussian distribution**

Next, we investigate the tuning of the standard deviation of a Gaussian distribution through VCMA [16,17]. VCMA enables the modulation of magnetic anisotropy energy in ferromagnetic materials via an applied electric field across an ultrathin insulating barrier. This electric field affects orbital hybridization at the ferromagnet/insulator interface, thereby modifying the perpendicular magnetic anisotropy (PMA), offering a promising mechanism for low-power, high-speed control of magnetic properties in spintronic devices. Fig. 3a illustrates the application of voltage across the MTJ stack. This voltage modulates the PMA: a positive voltage reduces the PMA, whereas a negative voltage enhances it. As illustrated in Fig. 3b, this modulation reshapes the energy landscape experienced by the DW. Micromagnetic simulation snapshots illustrate different DW configurations—straight (state ①) and tilted (states ② and ③)—corresponding to different positions in the energy landscape. In the high-PMA regime, the energy barrier between the straight and tilted DW states is substantial, rendering thermal excitations insufficient to drive transitions between the two states. This causes the DW to mainly remain in the straight state, resulting in enhanced stability and a reduced standard deviation in the TMR readout. Conversely, when PMA is reduced, the energy barrier is lowered, enabling thermal fluctuations to more easily drive transitions between states. This increases the DW's sensitivity to thermal noise, leading to a larger standard deviation in TMR readout. This behavior is validated through micromagnetic simulations, as shown in Extended Data Fig. 5. To experimentally confirm this effect, we center the mean position of the DW and use the same method used in Fig. 2f to collect 5,000 data points under different voltage conditions. As anticipated, the standard deviation of the TMR signal varies with the applied voltage, following the predicted trend. Fig. 3c quantifies the relationship between voltage and standard deviation, while Fig. 3d displays sampled Gaussian distributions of TMR at $V_{read} = -0.3V, 0.05V, 0.3V$.

**Bayesian neural network implementations by magnetic probabilistic computing platform**

To demonstrate the practical applicability of our proposed platform, we integrated it into a complete BNN framework for inference tasks. Fig. 4a shows the flow diagrams of the neural network which consists of convolutional layers, maxpooling layers, dropout layers, flatten layers, and dense layers. The training of BNNs is finished in Google Colab [37] using TensorFlow Probability [38]. The detailed implementation is described in the Methods section. The first five weights from the convolutional layer are extracted and plotted in Fig. 4b. As shown, these weights follow Gaussian distributions, each characterized by distinct mean values and standard deviations. We then map our Gaussian distributions from the MPC device to the ideal pre-trained neural network weight to build our physical BNN (see Methods). Fig. 4c(i) displays a test input image from the CIFAR-10 dataset, corresponding to the class "cat" (label 3). The prediction results are shown in Fig. 4c(ii), where green bars indicate correct predictions and red bars denote incorrect ones. The width of each bar represents the 95% confidence interval for the corresponding class. These results demonstrate that the proposed BNN not only performs classification but also provides meaningful uncertainty estimation. To further calibrate the physical BNN, we validate the accuracy vs. epochs and the final accuracy reaches 78.5%( Fig. 4d), which is comparable with the result from state-of-the-art [39]. We also conduct a recognition test and obtain the confusion matrix (Fig. 4e) which indicates the recognition probability. The diagonal element corresponds to the correct prediction rate. To further calibrate the reliability of our physical BNN, we test and plot the reliability diagram in Fig. 4f. The calibration curve aligns well with the perfect calibration, which indicates the neural network is neither overconfident nor underconfident.

We next present a benchmarking analysis comparing the performance of the MPC device to that of standard 28 nm CMOS technology, with a detailed discussion provided in the Supplementary Information. This benchmarking utilizes both directly measured experimental parameters and experimentally derived estimates to assess the device against standard CMOS across key performance metrics, including area efficiency, throughput, and energy consumption. The comparative results are summarized in Fig. 4g.

First, we provide an overview of the standard approach by which CMOS circuits generate tunable Gaussian-distributed random numbers for BNN implementations. In conventional CMOS architectures, this process typically involves three main steps: (1) generating base uniform random numbers, either through deterministic circuits such as linear feedback shift registers (LFSRs), which simulate randomness algorithmically, or through true random number generators (TRNGs) that exploit physical phenomena such as thermal noise, metastable behavior in flip-flops, or jitter in ring oscillators to produce genuinely random outputs; (2) transforming these uniform random numbers into Gaussian-distributed variables using mathematical techniques, most commonly the Central Limit Theorem (CLT) approach or the Box–Muller transformation; and (3) applying scaling and shifting operations to achieve the desired mean and variance, typically implemented with digital multiplier and adder circuits.

The benchmarking results reveal that our MPC device achieves a five-orders-of-magnitude improvement in area efficiency and a three-orders-of-magnitude improvement in energy efficiency compared to conventional CMOS implementations. This substantial advantage arises from the device's use of physical mechanisms—specifically, DW motion and VCMA—to achieve tunable Gaussian random number generation. In contrast, CMOS circuits require a large number of transistors to implement Box–Muller or CLT-based transformations, along with additional scaling and shifting operations. The benchmarking analysis further shows that the throughput of the MPC device is in the same order as CMOS circuits, primarily constrained by the DW motion speed and the time required to initially position the DW. However, with further materials optimization, the DW velocity can be significantly improved [25,26]. It is also important to note that the presented benchmarking is based on the time required to generate a single random number. In practical applications where N identical random numbers with the same distribution are needed (with N typically large), the time consumption in CMOS circuits scales approximately linearly with N. In contrast, in the MPC device, once the DW is set to the desired position, the only time per random number is the fast sampling time. The DW position setting time, which accounts for the majority of time delay, becomes negligible over repeated sampling. As a result, the effective overall throughput of the MPC device can surpass that of conventional CMOS circuits in large-scale applications.

According to the defined figure of merit (FOM), our device achieves an overall FOM improvement of seven orders of magnitude compared to conventional approaches.

$$FOM = \frac{1}{Area \times Time \times Energy}$$

## Conclusion

In conclusion, we present a physics-driven MPC platform that integrates DW motion, thermally induced stochasticity in DW position, VCMA, and TMR. This architecture enables efficient, device-level sampling of tunable Gaussian-distributed random numbers, paving the way for the practical deployment of physical probabilistic computing in AI applications. Through experimental validation on the CIFAR-10 classification task, we demonstrate the feasibility of implementing BNNs using our MPC hardware. Compared to standard 28 nm CMOS implementations, our approach achieves a seven-orders-of-magnitude improvement in the overall figure of merit, offering substantial gains in area, energy efficiency, and throughput. These results establish spintronic probabilistic computing as a viable technological foundation for next-generation trustworthy AI systems.

Looking ahead, the next step is to extend this platform toward integrated probabilistic processing units capable of executing entire stochastic inference pipelines in hardware. By scaling the MPC architecture into larger arrays and embedding it within compute-in-memory frameworks, more complex AI tasks can be supported, such as generative modeling. This advancement will further position spintronic probabilistic computing as a promising pathway toward the realization of reliable, trustworthy, and physically grounded AI systems.

**Acknowledgements**

The authors in University of California, Los Angeles acknowledge the support from the National Science Foundation (NSF) Award No. 1810163, No. 1611570, and No. 2328974; and the Army Research Office Multidisciplinary University Research Initiative (MURI) under grant numbers W911NF16-1-0472 and W911NF-19-S-0008.

## Author contributions

T.W. and B.D. designed, planned, and initiated studies. K.W. and Y.L. fabricated the devices. T.W. and B.D. conducted the MOKE and transport measurements and analyzed the data. T.W. performed the neural network training. K.L.W. supervised the project. T.W., B.D., and K.L.W. drafted the manuscript. All authors discussed the results and commented on the manuscript.

## Competing interests

The authors declare no competing interests.

## Data availability

All data are available in the main text or the supplementary materials upon request.

## Methods

### Math foundations of Bayesian neural network

In conventional neural networks, weights are optimized by simply maximizing the likelihood function, resulting in deterministic and fixed values for each parameter, as depicted in Fig. 1a(i). BNNs maintain a similar architectural framework but model the weights as probability distributions rather than point estimates, as illustrated in Fig. 1a(ii). The probabilistic feature enables BNNs to quantify uncertainty in predictions, thereby enhancing model robustness and reliability. The foundation of a BNN is rooted in Bayes' theorem, which is expressed as:

$$p(w|D) = \frac{p(D|w)p(w)}{p(D)}$$

In this equation, $w$ denotes the neural network weights. $p(w)$ represents the prior distribution over the weights, capturing prior beliefs about their values before observing any data. $D$ denotes the training dataset, and $p(D)$ is the marginal likelihood, serving as a normalization constant. The term $p(D|w)$ corresponds to the likelihood, indicating how well the model explains the observed data with weights $w$. $p(w|D)$ is the posterior distribution, which reflects the updated beliefs about the weights after incorporating the information provided by the data. To make predictions for a new input $x^*$, we need to integrate over all possible weights:

$$p(y^*|x^*, D) = \int p(y^*|x^*, w) \cdot p(w|D) \cdot dw$$

However, the posterior distribution $p(w|D)$ is generally intractable in neural networks due to the high dimensionality and nonlinearity of the parameter space.

In practice, Variational Inference (VI) [28-30] is commonly utilized to approximate the posterior with a simpler, parameterized distribution. VI introduces an alternative distribution $q(w; \phi)$, selected to be computationally tractable, and aims to make it as close as possible to the true posterior. In this context, $q(w; \phi)$ is typically chosen from a family of Gaussian distributions, where the network weights are modeled as Gaussian random variables, and $\phi$ denotes the means and standard deviations of the distributions. The objective is then to optimize

$\phi$ such that:

$$q(w;\phi) \approx p(w|D)$$

To quantify the difference between the selected Gaussian distribution and the true posterior, the KL divergence is introduced [31]. The KL divergence is a statistical measure that quantifies the discrepancy between a given probability distribution and a reference (or true) distribution. According to its definition, the goal is to minimize the following term:

$$KL(q(w;\phi)||p(w|D)) = \int q(w;\phi) \ln \frac{q(w;\phi)}{p(w|D)} dw$$

Minimizing the KL divergence is equivalent to maximizing the ELBO [32], which serves as the final objective function to be optimized during the training of a BNN.

$$\mathcal{L}(\phi) = \mathbb{E}_{q(w;\phi)}[\ln p(D|w)] - KL(q(w;\phi)||p(w))$$

During training, weights are first sampled as Gaussian distributions using the reparameterization method [33]. For each mini-batch, the network conducts a forward pass using sampled weights, evaluates the expected log-likelihood, and computes the corresponding KL divergence. The ELBO is optimized by iteratively updating the parameters of the weight distributions. This training procedure enables the BNN to simultaneously learn accurate predictive behavior and effectively capture model uncertainty.

**MOKE microscopy measurement**

Polar magneto-optic Kerr effect (MOKE) imaging was conducted using a custom-built wide-field MOKE microscope, offering a spatial resolution of 360 nm and a temporal resolution of 20 ms. This technique captures the magnetic signal along the out-of-plane (z) direction. An external magnetic field was applied using a GMW-5201 Helmholtz coil, powered by a Kepco BOP 5-20D supply. To begin, a large negative out-of-plane magnetic field was applied to fully saturate the sample, after which a background image was acquired. Subsequent images were obtained by subtracting this background, thereby enhancing magnetic contrast.

**Device design and fabrication**

The detailed design of the MPC device is presented in Extended Data Fig. 1. Extended Data Fig. 1a illustrates the mask layout. Each device consists of SOT current channels (Channels 1 and 2), MTJ reading channels (Channels 3 and 4), and two pairs of Oersted current channels (Channels 5 and 6, as well as Channels 7 and 8). SOT current injected through Channels 1 and 2 drives the DW movement along the magnetic stripe. The TMR signal is detected through Channels 3 and 4, which are connected respectively to the top and bottom electrodes of the MTJ. DWs are initialized using Oersted currents applied either through Channels 5 and 6 or through Channels 7 and 8. Extended Data Fig. 1b provides a zoomed-in view of the region marked by the red rectangle in Extended Data Fig. 1a. In this figure, the grey layer denotes the DW channel designed for DW propagation, the red layer represents the Ta contact layer that connects to the bottom electrode of the MTJ, and the black layer outlines the MTJ region. Extended Data Fig. 1c depicts the fabricated device along with the typical measurement setup, where SOT current is injected into the bottom Ta layer and TMR signal is measured across the MTJ stack by applying a reading voltage. Extended Data Figs. 1d and 1e illustrate cross-sectional views of the fabricated device along the x–z and y–z planes, respectively. MPC devices are fabricated using standard photolithography, dry etching, wet etching, and evaporation techniques. For detailed fabrication steps, refer to Extended Data Fig. 2.

## Domain generation, domain shifting, and TMR measurement

To initialize the domain state shown in Main Text Fig. 2, the MPC device was first saturated with a large negative out-of-plane magnetic field. Upon removal of the field, the device retained a negative out-of-plane magnetization. A current pulse generated by a Keithley 2636A Source-Meter was applied to the Oersted field channel to generate a magnetic domain with +z orientation (Extended Data Fig. 3a). The domain was then precisely repositioned to specific locations, enabling controlled targeting of various mean TMR values. Domain shifting was achieved through a sequence of SOT current pulses applied to the shifting channel using an Agilent 33250A arbitrary function generator (Extended Data Fig. 3b). Subsequent TMR measurements were performed with the Keithley 2636A Source-Meter. The typical experimental setup used for these measurements is illustrated in Extended Data Fig. 1c, demonstrating how a voltage was applied across the MTJ stack. All described measurements were conducted at room temperature.

## Gaussian distribution test

To verify that the data collected from the MPC device follows a Gaussian distribution, we utilized a quantile–quantile (Q–Q) plot [40] for validation and visualization. A Q–Q plot compares quantiles of the observed dataset with those from a theoretical Gaussian distribution. The procedure includes sorting experimental data in ascending order, calculating theoretical quantiles based on the Gaussian distribution corresponding to each data point, and plotting these sorted observed data quantiles (y-axis) against the theoretical quantiles (x-axis). If the observed data are normally distributed, the plotted points will align closely with the diagonal reference line. The ideal Gaussian Q–Q plot is illustrated in Extended Data Fig. 4a, demonstrating the expected alignment when both sample and theoretical data follow a Gaussian distribution. The Q–Q plot generated from data measured with our MPC device is presented in Extended Data Fig. 4b. Here, the experimental data points closely follow the reference line, confirming strong agreement with a Gaussian distribution.

**Micromagnetic simulation**

Due to the absence of direct imaging of DW fluctuations in the MPC device, we performed Mumax3 micromagnetic simulations to qualitatively investigate DW dynamics under varying PMA conditions. We adjusted the anisotropy constant to simulate the VCMA effect. The grid size is set as $400 \times 30 \times 1\ nm^3$ and cell size is set as $2 \times 2 \times 1\ nm^3$. The following material parameters are chosen for the simulation: exchange constant $A_{ex} = 10^{-11} J/m$, saturation magnetization $M_{sat} = 9 \times 10^5 A/m$, out-of-plane uniaxial anisotropy, interfacial DMI $3 \times 10^{-3} J/m^2$, temperature $T = 20K$, damping constant $\alpha = 0.015$. For low and high PMA cases, we use $K_{u1} = 8 \times 10^5 J/m^3$ and $K_{u2} = 1.2 \times 10^6 J/m^3$. The simulation result is shown in Extended Data Fig. 5. Different PMA conditions lead to varying DW stability, resulting in different TMR readout variations.

### BNN training

To evaluate the performance of the BNNs based on our MPC platform, we carried out the following experimental steps. Main text Fig. 4a illustrates the architecture of the neural network, comprising convolutional layers, max-pooling layers, dropout layers, flatten layers, and dense layers. Convolutional layers with configurations of 3×3×32, 3×3×64, and 3×3×128 are employed to extract image features such as edges and patterns. Max-pooling layers with a 2×2 window size are used to emphasize image features while simultaneously reducing image dimensions to lower computational costs. Dropout layers with a rate of 0.25 mitigate overfitting, thereby improving the network's generalization capabilities. One flatten and two dense layers (with 128 and 10 neurons each) constitute the network's final processing stages.

The BNN training was conducted on the Google Colab platform using TensorFlow Probability, where the network was trained for 250 epochs with the ADAM optimizer. During training, each weight in the network was modeled as a Gaussian distribution characterized by a mean and standard deviation. After training, these learned statistical parameters were mapped to experimentally accessible values from our devices.

To enable physical implementation, each trained mean and standard deviation was discretized to match the quantized levels achievable by the magnetic probabilistic hardware. Specifically, the mean values were mapped to discrete TMR levels that correspond to the DW position beneath the MTJ read head. The standard deviation was mapped to the experimentally measured thermal fluctuation range of the DW position at each control state, which manifests as variability in the TMR output. This mapping ensures that each learned Gaussian weight distribution in the trained model is approximated by a corresponding physical distribution generated by the device.

In effect, the trained virtual network was translated into a physically realizable implementation, where both the central value and spread of each weight distribution are emulated by the stochastic behavior of the MPC system. The performance of this mapped experimental network was then evaluated using the test dataset. Key metrics such as average recognition accuracy, confusion matrix, and reliability analysis were assessed and reported in Main Text Fig. 4.

**Validation accuracy:** It characterizes the overall predictive accuracy of the BNN on the validation dataset, assessing how well the model generalizes beyond training data. Our BNN reaches an overall accuracy of 78.5%, indicating a good prediction accuracy.

**Confusion matrix:** It provides a detailed breakdown of prediction performance by illustrating the distribution of true versus predicted classes. Specifically, it highlights the probability of classifying an input sample image into a particular category, facilitating a precise evaluation of classification accuracy and errors. The diagonal elements of the matrix represent correctly classified instances.

**Calibration curve:** It evaluates the reliability of the uncertainty estimates by plotting predicted probabilities against observed frequencies. A well-calibrated BNN yields a curve close to the ideal diagonal line, indicating accurate uncertainty quantification.

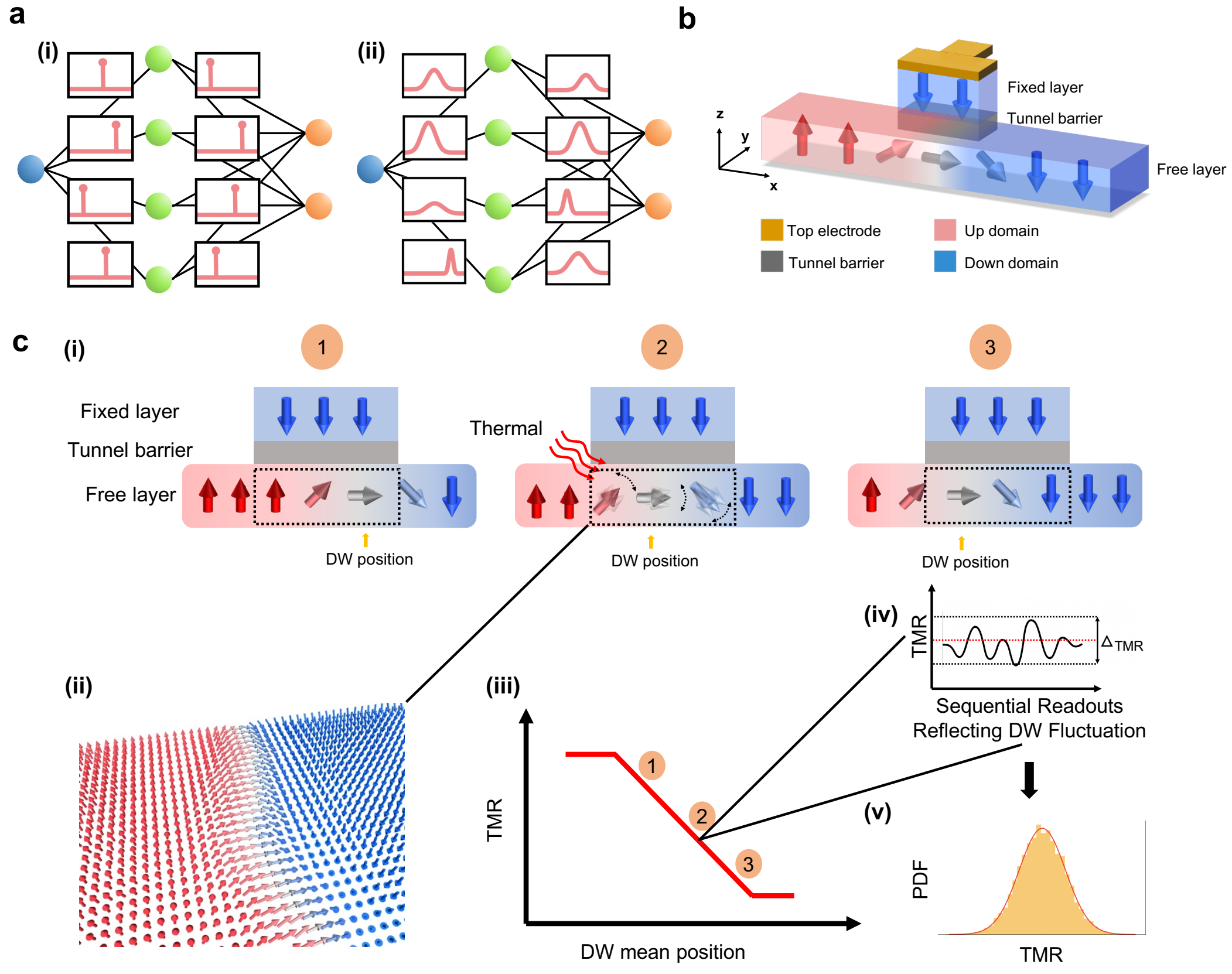

**Fig. 1 | Magnetic probabilistic computing platform and physical implementation of tunable probability. (a)** Structure of neural networks. (i) A point-estimate neural network, where each weight is represented by a single value. (ii) Structure of a BNN, where each weight is represented by a trainable Gaussian distribution. **(b)** Schematic of the MPC device. The device consists of three layers: the free layer, the tunnel barrier layer, and the fixed layer. Red (blue) arrows indicate +z (-z) magnetic domains. Grey arrow represents the in-plane (+x) magnetization direction, corresponding to the magnetic DW. The top electrode, connected to the fixed layer, functions as the readout electrode. A reading voltage is applied across the MTJ stack to measure the TMR signal. **(c)** Operating mechanism of the device. (i) DWs located at distinct locations (labeled ①, ②, and ③) along the free layer. The dashed rectangle indicates the region probed by the TMR read head. Thermal excitations depicted by red wavy arrows perturb the magnetic moments, causing stochasticity in the DW position. (ii) Micromagnetic simulation illustrating DW snapshots. (iii) Each DW position corresponds to a specific point on the TMR versus DW mean position curve. (iv) When the DW mean position is held at a fixed location (e.g., Position ②), repeated TMR readouts exhibit stochastic fluctuations. (v) The resulting TMR signal variations follow a Gaussian distribution, as illustrated by the probability density function (PDF) plot.

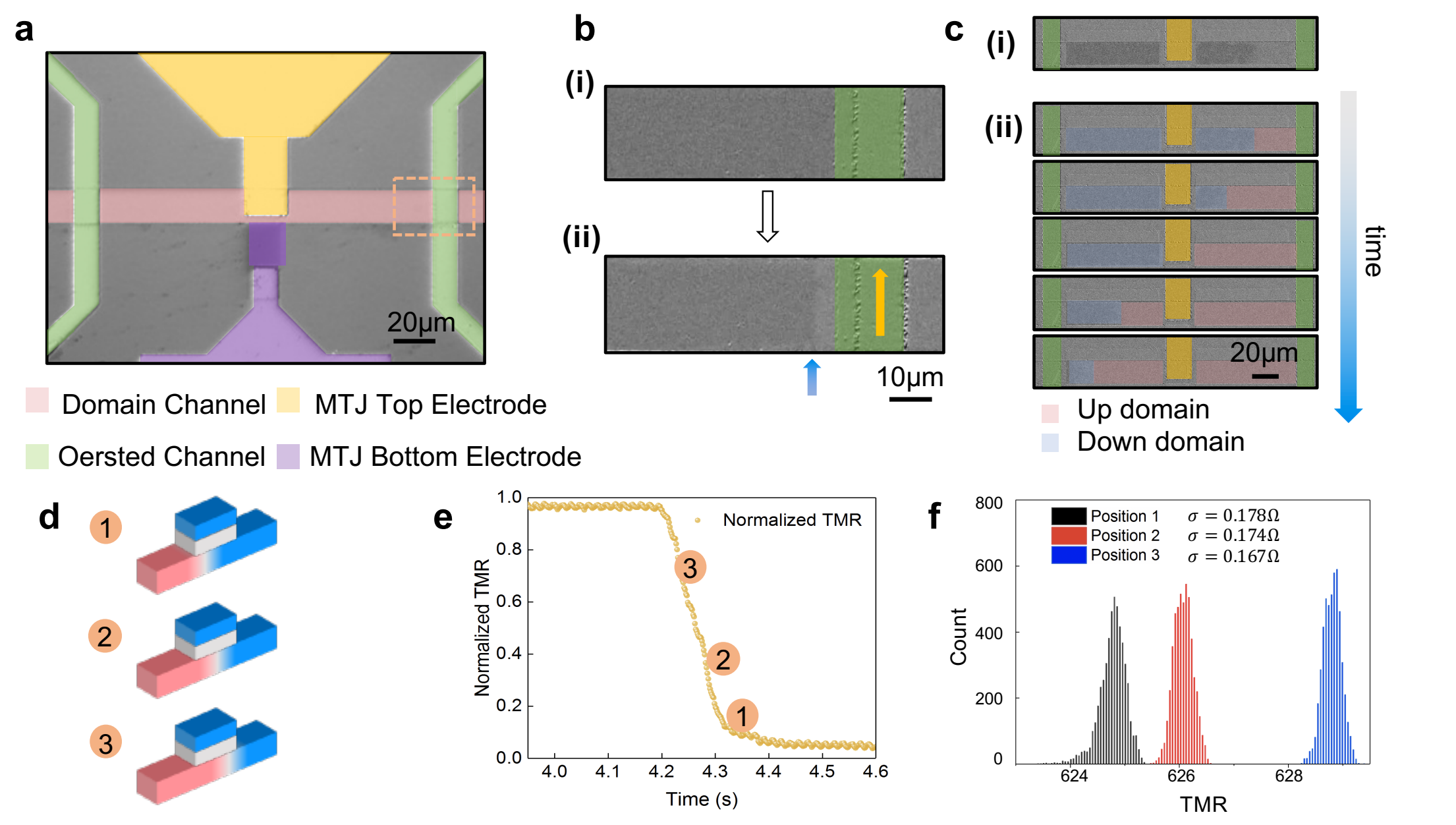

**Fig. 2 | Experimental realization of the magnetic probabilistic computing device. (a)** Microscope image of a fabricated MPC device. **(b)** Experimental demonstration of DW creation: (i) Zoomed-in image of the domain channel and Oersted channel, corresponding to the region within the orange dashed box in Fig. 2a, (ii) MOKE image showing the DW formation after applying a current pulse to the Oersted channel, as indicated by the yellow arrow. The MOKE image contrasts regions of different magnetization: areas with magnetization pointing up ($+M_Z$) appear white, while those with magnetization pointing down ($-M_Z$) appear black. The boundary between these regions, marked by the blue arrow, indicates the position of the domain wall. **(c)** Experimental realization of DW motion in the MPC device characterized by MOKE: (i) Snapshot of DW motion. (ii) Sequence of images showing SOT-driven DW motion from right to left. The up (down) domain region is shaded by red (blue) for better visualization. **(d)** Schematic illustration of the DW positioned at different locations beneath the MTJ. **(e)** Tuning of Gaussian mean by DW position: TMR characterization of the MPC device as the DW propagates through the MTJ. Different TMR values correspond to different DW positions. The location of labels ①, ②, and ③ presents the correspondence between the TMR signal and the DW's position in Fig. 2d, enabling the control of Gaussian mean. **(f)** Experimentally obtained Gaussian distribution: Sampling of Gaussian distribution TMR data with varying mean values and a fixed standard deviation. A total of 5,000 data points is collected for each DW position.

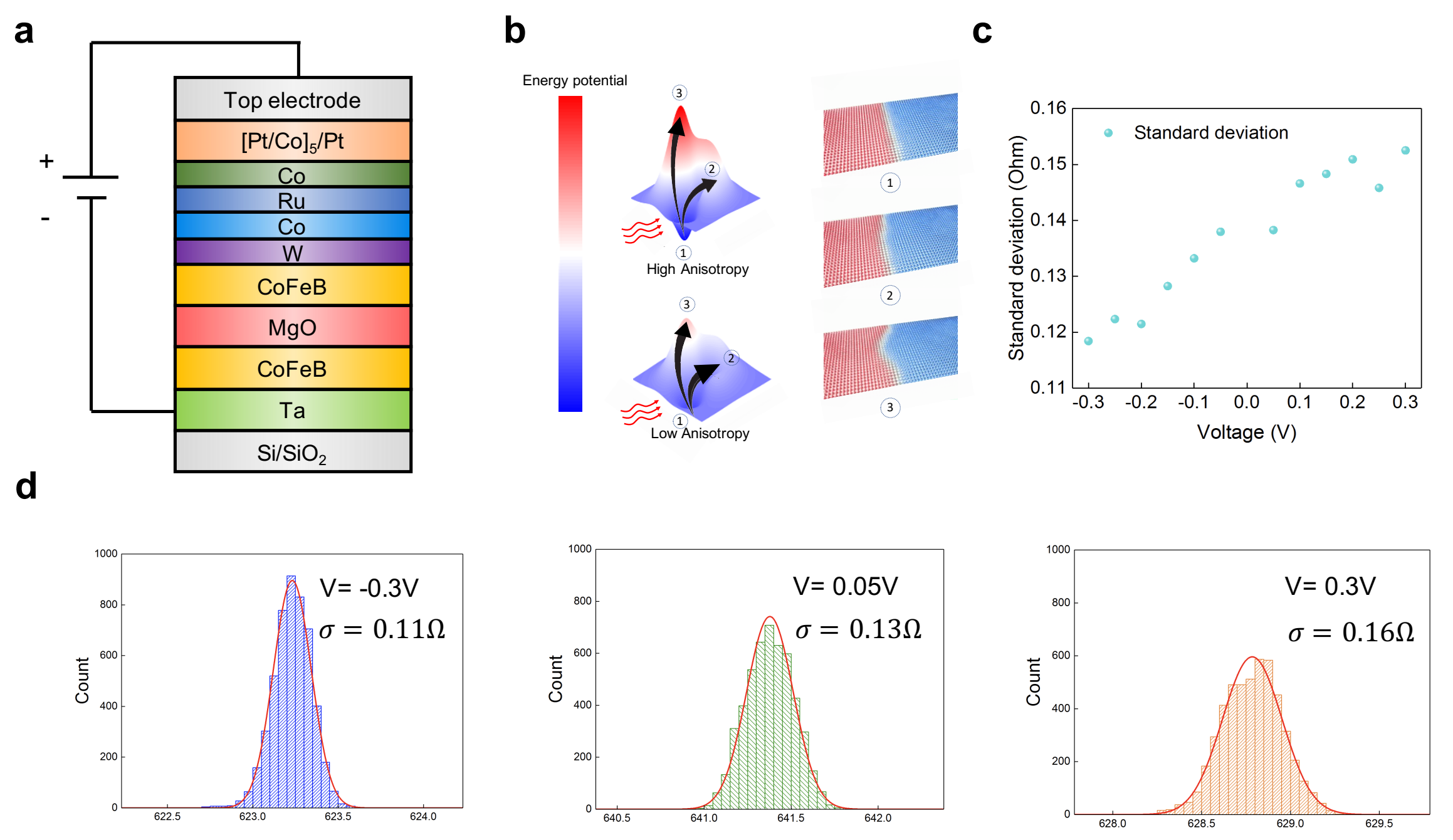

**Fig. 3 | Voltage-controlled standard deviation of Gaussian distribution. (a)** Materials stack. Applying a voltage across the MTJ modulates the perpendicular magnetic anisotropy (PMA): a positive (negative) voltage reduces (increases) the PMA. **(b)** Illustration of the DW energy landscape and corresponding DW configuration. Voltage-controlled magnetic anisotropy (VCMA) modulates the PMA and reshapes the energy landscape. Under thermal excitation, DWs in a low-PMA state are more likely to transition into an unstable (tilted) configuration due to a reduced energy barrier, leading to a larger standard deviation in the TMR readout. **(c)** Relationship between voltage and standard deviation. As the anisotropy changes, the DW's sensitivity to thermal fluctuations varies, resulting in different standard deviations in TMR. **(d)** Sampling of Gaussian distribution TMR data with varying standard deviations. A total of 5,000 data points is collected for each voltage condition (V = -0.3 V, 0.05 V, 0.3 V).

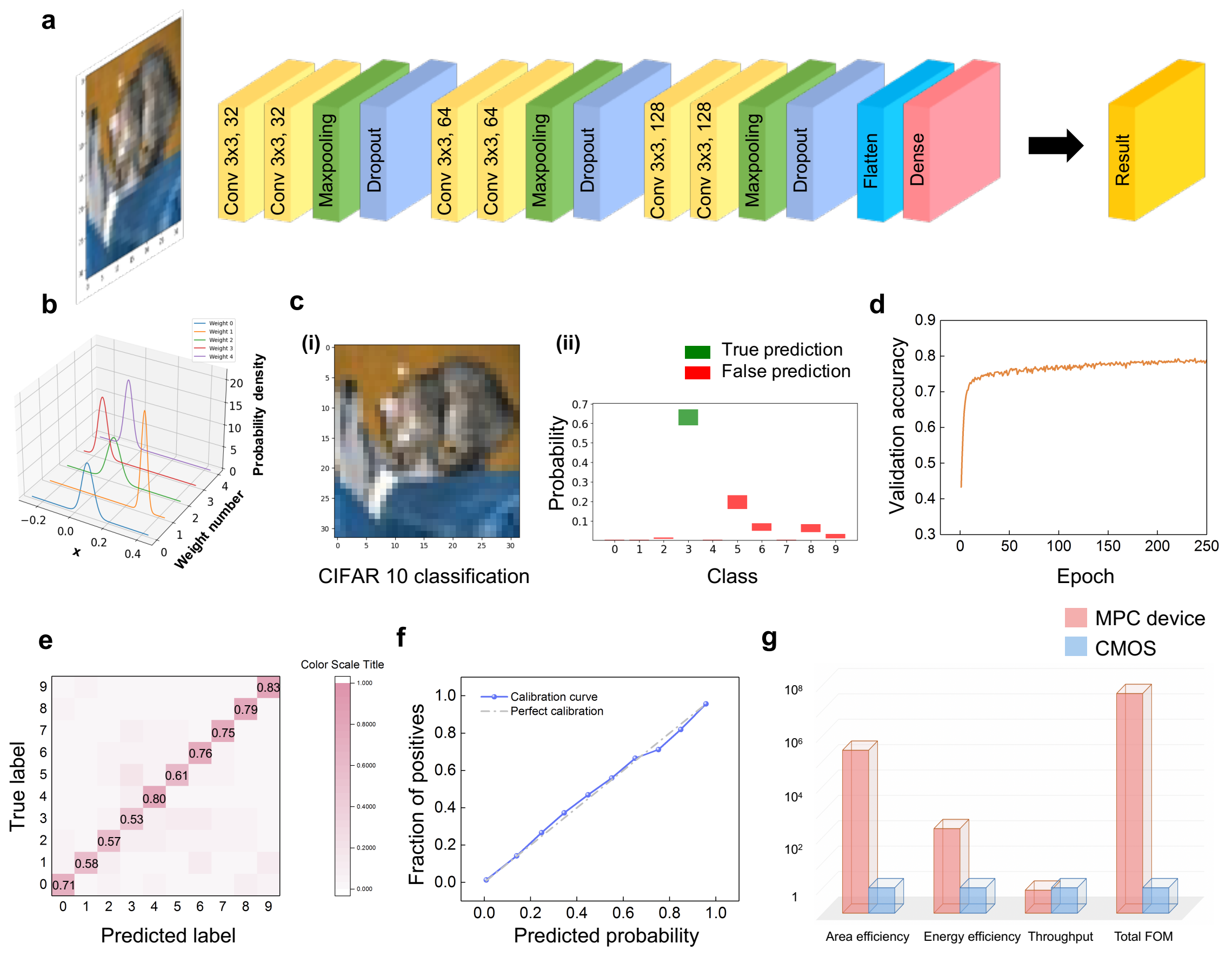

**Fig. 4 | Bayesian neural network implementations by magnetic probabilistic computing platform and benchmarking against CMOS. (a)** BNN flow diagram. **(b)** Examples of Gaussian distributions in the first convolutional layer. **(c)** BNN prediction result. (i) CIFAR-10 test image (cat, true label 3). (ii) BNN prediction on a CIFAR-10 test image. Green bar indicates correct predictions, while red bar indicates incorrect predictions. The width of the color bar represents the 95% confidence interval. **(d)** Validation accuracy vs. epochs. The final validation accuracy reaches 78.5% after 250 training epochs. **(e)** Confusion matrix. The y-axis represents the true labels, while the x-axis represents the predicted labels. The color scale indicates recognition probability, with the diagonal showing correct prediction rates. **(f)** Calibration curve. The curve closely aligns with perfect calibration, indicating neither overconfidence nor underconfidence. **(g)** Benchmarking between MPC device and 28nm CMOS technology in a logarithmic scale. MPC device achieves 7 orders of magnitude in FOM.

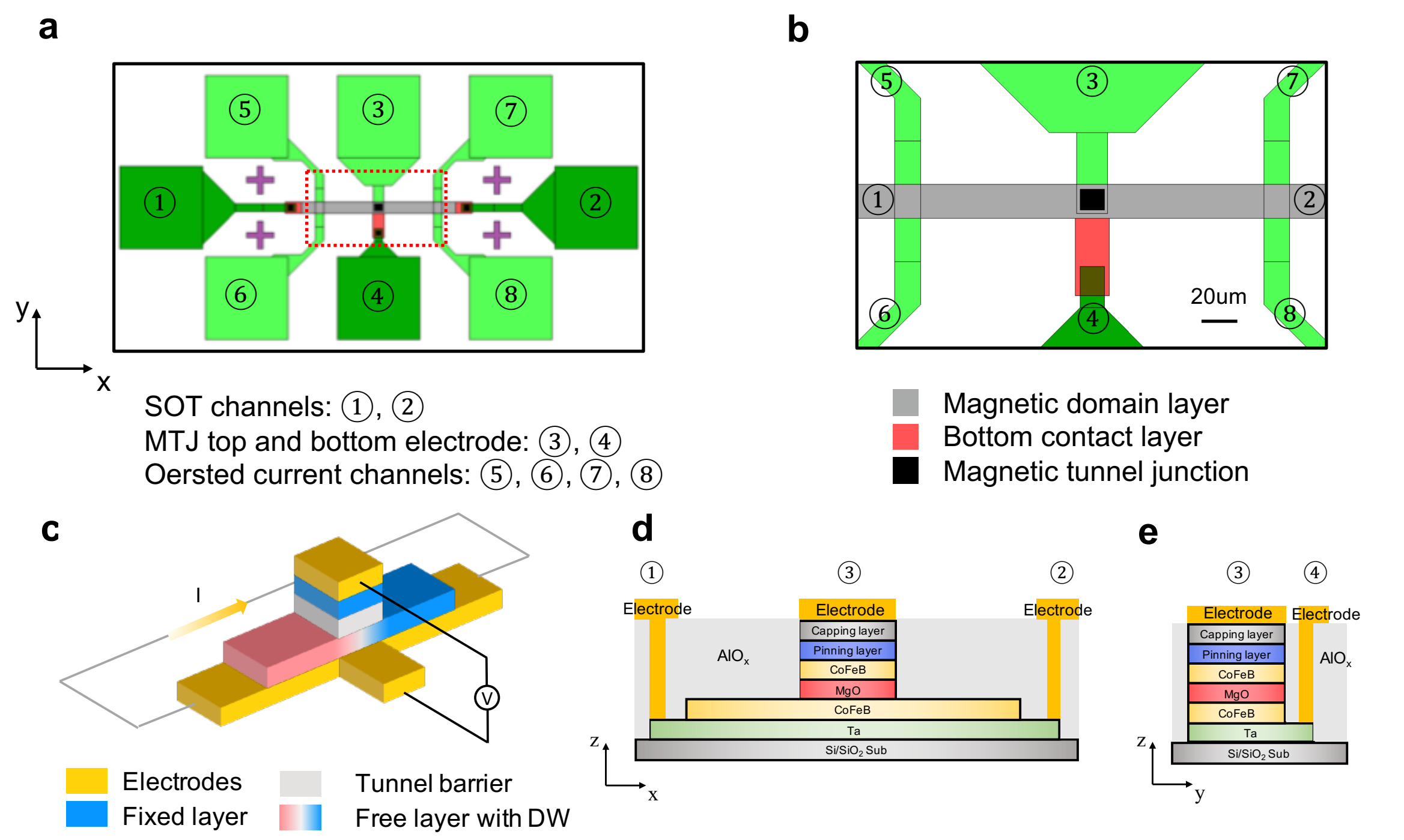

**Extended Data Fig. 1 | Mask design. (a)** Mask layout of MPC device. Spin-orbit torque (SOT) current is injected through Channels 1 and 2, which drives the DW along the magnetic stripe. The TMR signal is read out via Channels 3 and 4, which connect to the top and bottom electrodes of the MTJ respectively. DWs are initialized by Oersted currents applied through Channels 5 and 6, alternatively through Channels 7 and 8. **(b)** Detailed view of the MPC device. The grey layer represents the DW channel for DW motion. The red layer illustrates the bottom Ta contact layer, connecting the free layer of the MTJ. The black layer outlines the MTJ area, highlighting the location of MTJs. **(c)** Illustration of fabricated device and measurement setup. SOT current is injected in the bottom Ta layer. The TMR signal is measured across the MTJ stack by applying a vertical reading voltage. **(d)** Cross-section in the x-z plane of the fabricated device. **(e)** Cross-section in the y-z plane of the fabricated device.

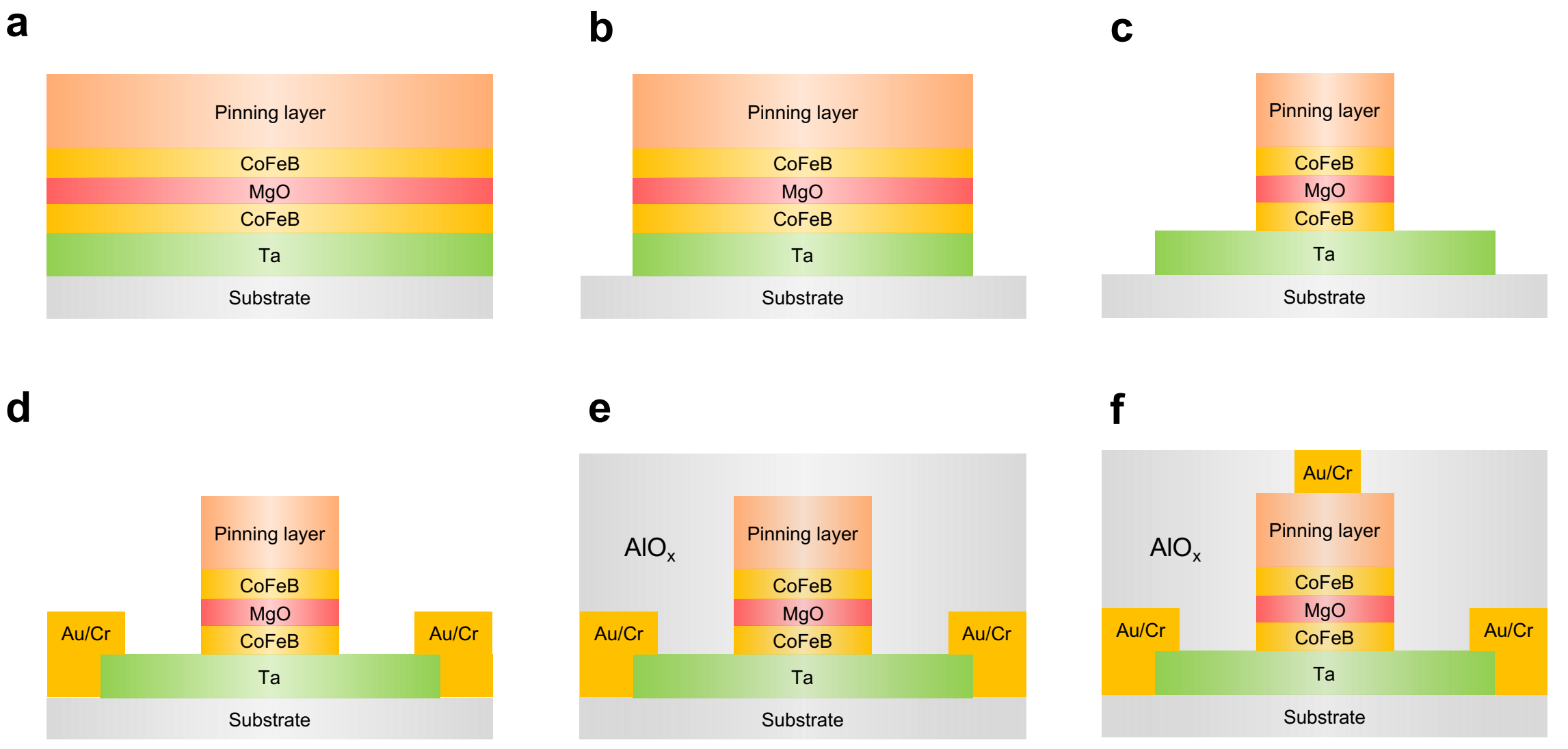

**Extended Data Fig. 2 | Device fabrication. (a)** Simplified MTJ stack. **(b)** Photolithography and etching for device shape. **(c)** Photolithography and etching for MTJ pillar. **(d)** Photolithography and evaporation for Au/Cr MTJ Bottom electrodes. **(e)** $AlO_x$ deposition by Atomic Layer Deposition (ALD). **(f)** Photolithography, etching, and evaporator for Au/Cr MTJ top electrode.

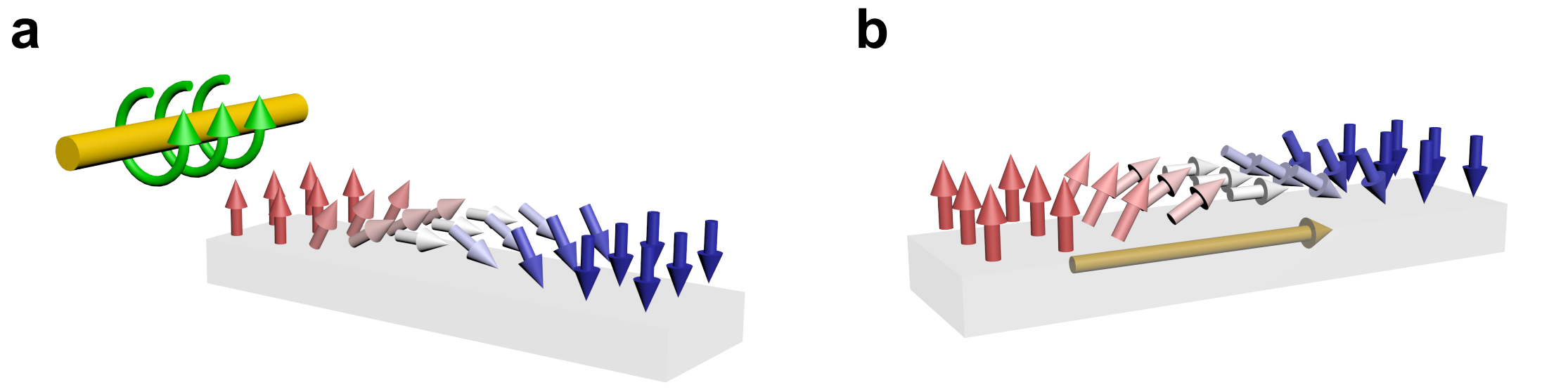

**Extended Data Fig. 3 | Mechanism of domain creation and DW motion. (a)** Schematic of domain generation. The yellow tube represents the conducting stripline, while the green arrows indicate the Oersted field generated by the current. Current pulses induce the Oersted field, which facilitates the nucleation of magnetic domains. **(b)** Schematic of DW motion. Yellow arrow represents the SOT current.

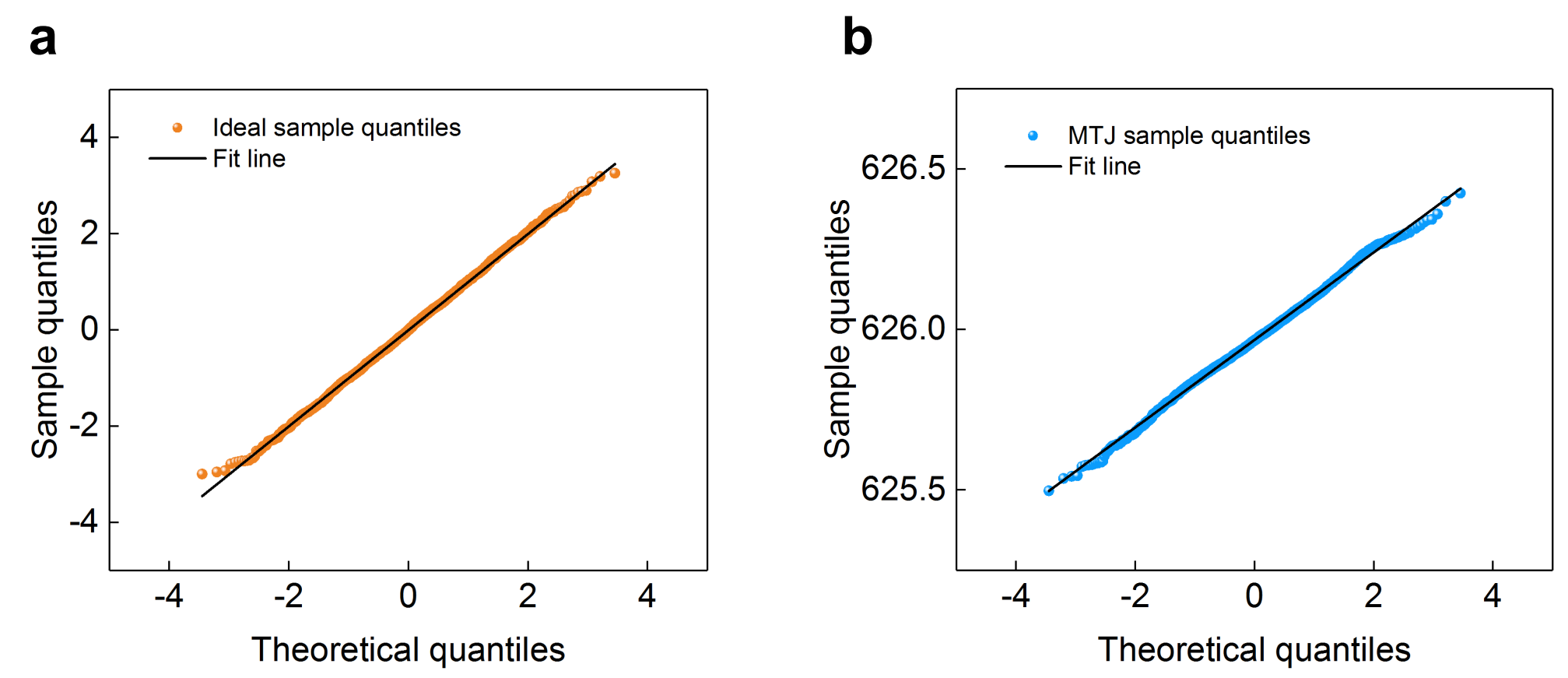

**Extended Data Fig. 4 | Gaussian distribution analysis. (a)** Quantile–Quantile plot validating that the data generated by Python follows an ideal Gaussian distribution. **(b)** Quantile-Quantile plot evaluating the Gaussian distribution of data produced by the MPC device. The data closely aligns with the reference line, indicating an excellent Gaussian fit.

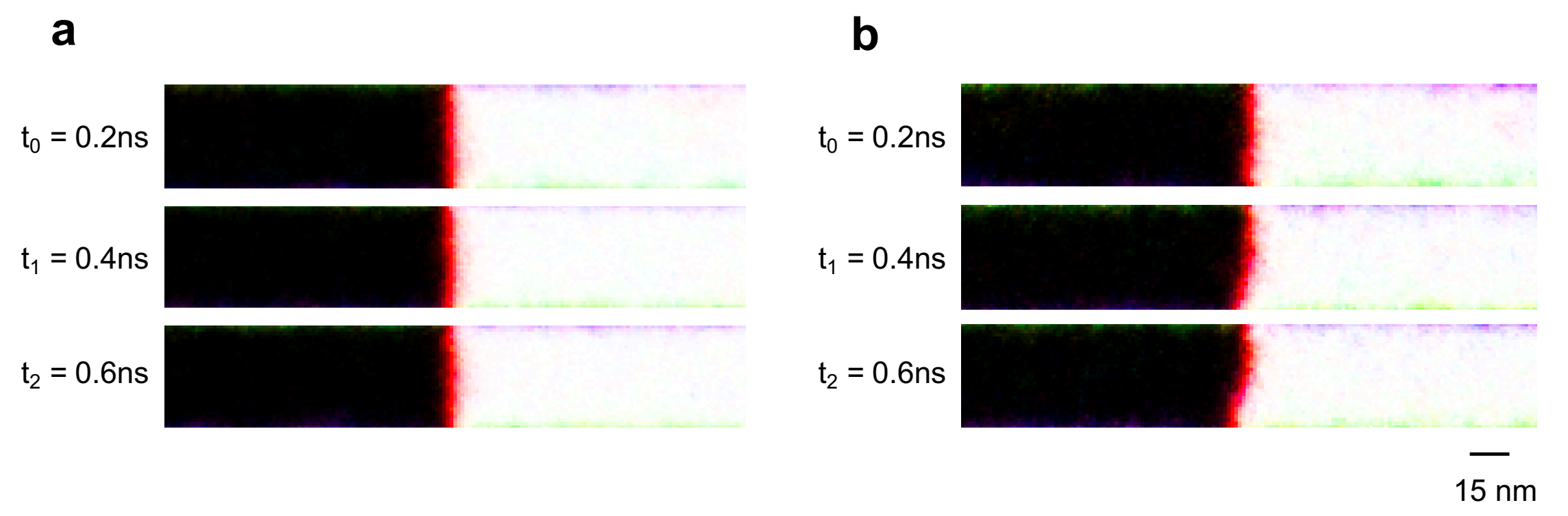

**Extended Data Fig. 5 | Micromagnetic simulation for high PMA and low PMA case comparison. (a)** DW time evolution images exhibit fluctuation under thermal excitation in the high perpendicular magnetic anisotropy (PMA) state. The DW remains stable with minimal tilting due to strong PMA. **(b)** DW exhibits increased tilting under thermal excitation in the low PMA state, indicating reduced stability.

**Reference**

1. Kaur, D., Uslu, S., Rittichier, K.J. and Durresi, A., 2022. Trustworthy artificial intelligence: a review. *ACM computing surveys (CSUR)*, *55*(2), pp.1-38.
2. Thiebes, S., Lins, S. and Sunyaev, A., 2021. Trustworthy artificial intelligence. *Electronic Markets*, *31*, pp.447-464
3. Mohebali, B., Tahmassebi, A., Meyer-Baese, A. and Gandomi, A.H., 2020. Probabilistic neural networks: a brief overview of theory, implementation, and application. *Handbook of probabilistic models*, pp.347-367.
4. Ngartera, L., Issaka, M.A. and Nadarajah, S., 2024. Application of Bayesian Neural Networks in Healthcare: Three Case Studies. *Machine Learning and Knowledge Extraction*, *6*(4), pp.2639-2658.
5. Loquercio, A., Segu, M. and Scaramuzza, D., 2020. A general framework for uncertainty estimation in deep learning. *IEEE Robotics and Automation Letters*, *5*(2), pp.3153-3160.
6. Zhou, T., Han, T. and Droguett, E.L., 2022. Towards trustworthy machine fault diagnosis: A probabilistic Bayesian deep learning framework. *Reliability Engineering & System Safety*, *224*, p.108525.
7. Huang, P., Gu, Y., Fu, C., Lu, J., Zhu, Y., Chen, R., Hu, Y., Ding, Y., Zhang, H., Lu, S. and Peng, S., 2023. SOT-MRAM-Enabled Probabilistic Binary Neural Networks for Noise-Tolerant and Fast Training. arXiv preprint arXiv:2309.07789.

8. Yu, G., Upadhyaya, P., Wong, K.L., Jiang, W., Alzate, J.G., Tang, J., Amiri, P.K. and Wang, K.L., 2014. Magnetization switching through spin-Hall-effect-induced chiral domain wall propagation. *Physical Review B*, *89*(10), p.104421.
9. Parkin, S.S., Hayashi, M. and Thomas, L., 2008. Magnetic domain-wall racetrack memory. *Science*, *320*(5873), pp.190-194.
10. Fert, A., Reyren, N. and Cros, V., 2017. Magnetic skyrmions: advances in physics and potential applications. *Nature Reviews Materials*, *2*(7), pp.1-15.
11. Nagaosa, N. and Tokura, Y., 2013. Topological properties and dynamics of magnetic skyrmions. *Nature Nanotechnology*, *8*(12), pp.899-911.
12. Jiang, W., Upadhyaya, P., Zhang, W., Yu, G., Jungfleisch, M.B., Fradin, F.Y., Pearson, J.E., Tserkovnyak, Y., Wang, K.L., Heinonen, O. and Te Velthuis, S.G., 2015. Blowing magnetic skyrmion bubbles. *Science*, *349*(6245), pp.283-286.
13. Thomas, L., Hayashi, M., Jiang, X., Moriya, R., Rettner, C. and Parkin, S., 2007. Resonant amplification of magnetic domain-wall motion by a train of current pulses. *Science*, *315*(5818), pp.1553-1556.
14. Dai, B., Wu, D., Razavi, S.A., Xu, S., He, H., Shu, Q., Jackson, M., Mahfouzi, F., Huang, H., Pan, Q. and Cheng, Y., 2023. Electric field manipulation of spin chirality and skyrmion dynamic. *Science Advances*, *9*(7), p.eade6836.
15. Yu, G., Upadhyaya, P., Shao, Q., Wu, H., Yin, G., Li, X., He, C., Jiang, W., Han, X., Amiri, P.K. and Wang, K.L., 2017. Room-temperature skyrmion shift device for memory application.*Nano letters*, *17*(1), pp.261-268.

16. Dai, B., Jackson, M., Cheng, Y., He, H., Shu, Q., Huang, H., Tai, L. and Wang, K., 2022. Review of voltage-controlled magnetic anisotropy and magnetic insulator. *Journal of Magnetism and Magnetic Materials*, *563*, p.169924.
17. Nozaki, T., Yamamoto, T., Miwa, S., Tsujikawa, M., Shirai, M., Yuasa, S. and Suzuki, Y., 2019. Recent progress in the voltage-controlled magnetic anisotropy effect and the challenges faced in developing voltage-torque MRAM. *Micromachines*, *10*(5), p.327.
18. Moodera, J.S., Kinder, L.R., Wong, T.M. and Meservey, R., 1995. Large magnetoresistance at room temperature in ferromagnetic thin film tunnel junctions. *Physical review letters*, *74*(16), p.3273.
19. Tsymbal, E.Y., Mryasov, O.N. and LeClair, P.R., 2003. Spin-dependent tunnelling in magnetic tunnel junctions. *Journal of Physics: Condensed Matter*, *15*(4), p.R109.
20. Yang, K., Malhotra, A., Lu, S. and Sengupta, A., 2020. All-spin Bayesian neural networks. *IEEE Transactions on Electron Devices*, *67*(3), pp.1340-1347.
21. Liu, S., Xiao, T.P., Kwon, J., Debusschere, B.J., Agarwal, S., Incorvia, J.A.C. and Bennett, C.H., 2022. Bayesian neural networks using magnetic tunnel junction-based probabilistic in-memory computing. *Frontiers in Nanotechnology*, *4*, p.1021943.
22. Ahmed, S.T., Danouchi, K., Hefenbrock, M., Prenat, G., Anghel, L. and Tahoori, M.B., 2023, April. Scalable spintronics-based bayesian neural network for uncertainty estimation. In *2023 Design, Automation & Test in Europe Conference & Exhibition (DATE)* (pp. 1-6). IEEE.

23. Gu, Y., Huang, P., Chen, T., Fu, C., Chen, A., Peng, S., Zhang, X. and Kou, X., 2024. A noise-tolerant, resource-saving probabilistic binary neural network implemented by the SOT-MRAM compute-in-memory system. arXiv preprint arXiv:2403.19374.

24. Wiesendanger, R., 2016. Nanoscale magnetic skyrmions in metallic films and multilayers: a new twist for spintronics. Nature Reviews Materials, 1(7), pp.1-11.

25. Kumar, D., Chung, H.J., Chan, J., Jin, T., Lim, S.T., Parkin, S.S., Sbiaa, R. and Piramanayagam, S.N., 2023. Ultralow energy domain wall device for spin-based neuromorphic computing. ACS nano, 17(7), pp.6261-6274.

26. Caretta, L., Oh, S.H., Fakhrul, T., Lee, D.K., Lee, B.H., Kim, S.K., Ross, C.A., Lee, K.J. and Beach, G.S., 2020. Relativistic kinematics of a magnetic soliton. Science, 370(6523), pp.1438-1442.

27. Guo, Z., Yin, J., Bai, Y., Zhu, D., Shi, K., Wang, G., Cao, K. and Zhao, W., 2021. Spintronics for energy-efficient computing: An overview and outlook. Proceedings of the IEEE, 109(8), pp.1398-1417.

28. Graves, A., 2011. Practical variational inference for neural networks. *Advances in neural information processing systems*, *24*.

29. Gal, Y. and Ghahramani, Z., 2015. Bayesian convolutional neural networks with Bernoulli approximate variational inference. *arXiv preprint arXiv:1506.02158*.

30. A Alemi, A.A., Morningstar, W.R., Poole, B., Fischer, I. and Dillon, J.V., 2020. Vib is half bayes. *arXiv preprint arXiv:2011.08711*.

31. Hershey, J.R. and Olsen, P.A., 2007, April. Approximating the Kullback Leibler divergence between Gaussian mixture models. In *2007 IEEE International*

*Conference on Acoustics, Speech and Signal Processing-ICASSP'07* (Vol. 4, pp. IV-317). IEEE.

32. Hoffman, M.D. and Johnson, M.J., 2016, December. Elbo surgery: yet another way to carve up the variational evidence lower bound. In *Workshop in advances in approximate bayesian inference, NIPS* (Vol. 1, No. 2).
33. Kingma, D.P. and Welling, M., 2013. *Auto-encoding variational bayes* [online]
34. Vansteenkiste, Arne, et al. "The design and verification of MuMax3." AIP advances 4.10 (2014).
35. Guo, Z., Yin, J., Bai, Y., Zhu, D., Shi, K., Wang, G., Cao, K. and Zhao, W., 2021. Spintronics for energy-efficient computing: An overview and outlook. *Proceedings of the IEEE*, *109*(8), pp.1398- 1417.
36. Liu, L., Pai, C.F., Li, Y., Tseng, H.W., Ralph, D.C. and Buhrman, R.A., 2012. Spin-torque switching with the giant spin Hall effect of tantalum. *Science*, *336*(6081), pp.555-558.
37. Bisong, E., 2019. Google colaboratory. In *Building machine learning and deep learning models on google cloud platform: a comprehensive guide for beginners* (pp. 59-64). Berkeley, CA: Apress.
38. Dillon, J.V., Langmore, I., Tran, D., Brevdo, E., Vasudevan, S., Moore, D., Patton, B., Alemi, A., Hoffman, M. and Saurous, R.A., 2017. Tensorflow distributions. *arXiv preprint arXiv:1711.10604*.

39. Shridhar, K., Laumann, F. and Liwicki, M., 2019. A comprehensive guide to bayesian convolutional neural network with variational inference. *arXiv preprint arXiv:1901.02731*.

40. Wilk, M.B. and Gnanadesikan, R., 1968. Probability plotting methods for the analysis for the analysis of data. *Biometrika*, *55*(1), pp.1-17.

## Supplementary Information of

# Spintronic Bayesian Hardware Driven by Stochastic Magnetic Domain Wall Dynamics

**Authors**

Tianyi Wang[1,*,†], Bingqian Dai[1,*,†], Kin Wong[1], Yaochen Li[1], Yang Cheng[1], Qingyuan Shu[1], Haoran He[1], Puyang Huang[1] , Hanshen Huang[1] and Kang L. Wang[1,†]

**Affiliations**

*[1]Department of Electrical and Computer Engineering, University of California, Los Angeles, California 90095, United States*

*These authors contributed equally to this work.

Corresponding author. E-mail: †tianyiwang0220@g.ucla.edu, †bdai@g.ucla.edu, †wang@ee.ucla.edu

We present a detailed area, throughput, and energy analysis of our magnetic probabilistic computing (MPC) device versus standard 28nm CMOS. This benchmarking utilizes both directly measured experimental parameters and experimentally derived estimates to assess the device against standard CMOS across key performance metrics, including area efficiency, throughput, and energy consumption. The comparative results are summarized in Main text Figure. 4g.

**Estimation for magnetic probabilistic computing device**

The area of the BNN MTJ device is primarily constrained by fabrication limits and the stability of magnetic domain wall. The MTJ width is set by the fabrication resolution, which in this work is assumed to be 30 nm according to TSMC 28nm technology node. The device length is determined by the accuracy required for generating Gaussian random numbers. To achieve 8-bit Gaussian random number generation, precise control over domain wall displacement is necessary. Previous studies have demonstrated sub-nanometer tunability of domain wall positions [1]; here, we conservatively assume a tunable step size of 1 nm. This yields an active MTJ pillar length of 256 nm (corresponding to $2^8$ steps). Additionally, to account for domain wall nucleation via the Oersted field, we include an extra 75 nm region for domain wall generation and stabilization. Therefore, the total device area is calculated as

$$Area = W_{MTJ} \times (L_{MTJ} + 75nm) = 30nm \times (256nm + 75nm) = 9.93 \times 10^{-3} \mu m^2$$

The total time latency arises from two main contributions: the domain wall motion time and the sampling time. Here we ignore the time needed for Oersted field-driven domain wall generation because such process is only required during the initialization. The domain wall motion time is governed by the domain wall velocity, reported in prior work to be approximately 25 m/s [2]. Accordingly, the time required to position the domain wall under the MTJ read head is estimated as:

$$t_{DWM} = \frac{L_{MTJ}}{v_{DW}} = \frac{256nm}{25m/s} = 10.24ns$$

The sampling time is determined by the RC time constant of the readout circuit, where we assume an MTJ resistance of 100 kΩ and a typical sampling capacitance of 1 fF, yielding:

$$\tau = R_{MTJ} \times C_{sample} = 0.1ns$$

Thus, the total estimated time latency per operation is:

$$t_{total} = t_{read} + t_{DWM} = 10.34ns$$

The calculated latency corresponds to the time required to generate a single Gaussian random number. However, when sampling a large number of Gaussian random numbers, the domain wall can be held at the mean position, and only the sampling time contributes to the overall latency. This significantly reduces the effective time delay per sample.

The total energy consumption consists primarily of two components: the energy required for domain wall shifting and the energy required for readout. To estimate the domain wall shifting energy, we use the following experimental parameters: a current density $J_{SOT} = 1.1 \times 10^{12} A/m^2$, Resistivity $\rho = 9.4 \times 10^{-7}$ m · Ω. Shift current can thus be calculated.

$$I_{shift} = J_{SOT} \times thickness \times W_{MTJ} = 1.65 \times 10^{-4} A$$

The resistance of the strip can be calculated as

$$R_{strip} = \rho \times \frac{L_{MTJ} + 75nm}{W_{MTJ} \times thickness} = 2.07k\Omega$$

The shift energy can be calculated as

$$E_{shift} = {I_{shift}}^2 \times R_{strip} \times t_{DWM} = 5.77 \times 10^{-13} J$$

We take the $V_{read}$ to be 1V, then the reading energy can be calculated as

$$E_{read} = \frac{{V_{read}}^2}{R_{MTJ}} \times t_{read} = 1 \times 10^{-15} J$$

The energy for charging the sampling capacitors can be calculated as

$$E_{cap} = C_{sample}{V_{read}}^2 = 1 \times 10^{-15} J$$

The total energy is

$$E_{total} = E_{shift} + E_{read} + E_{cap} = 0.579pJ$$

The parameters for estimation are listed in the table.

| Parameters | Definition | Value |
|---|---|---|
| $v_{DW}$ | Domain wall max velocity | $25m/s$ |
| $\rho$ | Strip resistivity | $9.4 \times 10^{-7}\ \mathrm{m} \cdot \Omega$ |
| $J_{SOT}$ | Shift current density | $1.1 \times 10^{12}\ A/m^2$ |
| $thickness$ | Strip thickness | $5\ nm$ |
| $R_{MTJ}$ | Resistance of MTJ | $100\ k\Omega$ |
| $L_{MTJ}$ | Length of MTJ | $256\ nm$ |
| $W_{MTJ}$ | Width of MTJ pillar | $30\ nm$ |
| $C_{sample}$ | Sample capacitor | $1\ fF$ |

**Estimation for CMOS circuits**

The ability to generate Gaussian-distributed random numbers with tunable mean and variance is critical in the Bayesian neural network implementation. In CMOS circuits, this process typically involves three major steps: (1) generating base uniform random numbers, (2) transforming these uniform numbers into Gaussian-distributed variables through mathematical techniques, (3) applying scaling and shifting operations to achieve the desired mean and variance. This section systematically describes the standard methods employed in each step to realize tunable Gaussian distributions in digital CMOS implementations.

Step 1: Several methods exist for generating uniformly distributed random numbers in CMOS circuits. Deterministic circuits, such as Linear Feedback Shift Registers (LFSRs), simulate randomness through algorithmic means. LFSRs, typically constructed from flip-flops and XOR gates, produce sequences that appear random but are in fact deterministic, categorizing them as pseudo-random number generators (PRNGs). In addition to deterministic approaches, true random number generators (TRNGs) leverage physical sources of randomness. TRNG circuits utilize phenomena such as thermal noise, metastable behavior in flip-flops, or jitter in ring oscillators to produce genuinely random outputs. These methods can create uniformly distributed random variables for further use.

Step 2: Two primary methods are commonly employed for generating Gaussian-distributed random variables: the Central Limit Theorem (CLT) and the Box–Muller transformation. The CLT states that the normalized sum (or mean) of a sufficiently large number of independent and identically distributed random variables converges to a standard normal distribution. However, due to its inherent reliance on sampling, the CLT-based method may exhibit poor accuracy in the distribution tails when the sample size is limited. To avoid such limitations and ensure high fidelity in the generated distribution, Box–Muller transformation is often adopted, which produces truly Gaussian-distributed random variables without requiring extensive sampling.

The Box–Muller transform, originally proposed by George Edward Pelham Box and Mervin Edgar Muller, is a mathematical technique for generating pairs of independent, standard normally distributed random numbers from uniformly distributed random numbers. Given two independent samples $x_1$ and $x_2$ drawn from a uniform distribution over the interval $(0, 1)$, the transformation is defined as follows:

$$z_1 = \sqrt{-2\ln x_1}\cos(2\pi x_2)$$

$$z_2 = \sqrt{-2\ln x_1}\sin(2\pi x_2)$$

Two samples $x_1$ and $x_2$ drawn from the uniform distribution over the interval $(0, 1)$, are transformed into $z_1$ and $z_2$, two independent standard normal random variables with zero mean and unit variance.

Step 3: In digital CMOS circuits, once a Gaussian-distributed random variable $z_0$ with zero mean and unit variance is generated, it is scaled to a target mean μ and standard deviation σ through a simple linear transformation:

$$z_{scaled} = \sigma z_0 + \mu$$

This scaling operation is typically implemented using digital arithmetic blocks. The standard normal variable is first multiplied by the desired standard deviation using a digital multiplier. The resulting scaled value is then offset by the target mean through a digital adder circuit, producing the final Gaussian-distributed output.

***Energy cost:***

CMOS-based random number generators typically consume approximately 2 pJ per generated bit. For instance, a digital TRNG implemented in CMOS was reported to achieve an energy consumption of 2.58 pJ per bit [3]. To further generate Gaussian-distributed random numbers from uniform inputs, Box–Muller transformation involves several computational stages: logarithm calculation, square root calculation, trigonometric functions, and multiplications. Logarithm and square root functions, typically realized using lookup tables or iterative methods, are computationally intensive and can consume approximately 100–200 pJ per operation. Trigonometric functions are often implemented using CORDIC algorithms, requiring 10–20 pJ per sine or cosine evaluation. [4, 5] Multiplication, when performed using an 8×8-bit multiplier in 28nm CMOS, generally consumes around 2 pJ per operation [6]. To scale the output to arbitrary mean and variance, the final stage includes an additional 8-bit multiplication and addition. This step adds approximately 3 pJ per sample, based on typical energy costs for multipliers (~2 pJ) and adders (<1 pJ).

Considering all stages—including uniform RNG generation, Box–Muller transformation, and final scaling—the total energy required to generate a single 8-bit Gaussian random number in 28nm CMOS is estimated to be in the range of 119–229 pJ.

***Time latency:***

Uniform random number generators (RNGs), typically implemented using linear feedback shift registers (LFSRs) or similar pseudo-random number circuits can produce output within **1–2 ns**. In prior work, a latency of **2.65 ns** has been reported for a CMOS RNG [7]. The latency for computing the Box Muller transformation depends heavily on the underlying implementation. If lookup tables (LUTs) are used, these functions can be completed in a single clock cycle. In contrast, iterative or CORDIC-based implementations typically require multiple clock cycles [8], with reported latencies ranging from **1 to 10 cycles** depending on pipeline depth, target precision, and clock speed. The multipliers used in the Box–Muller stage are generally fully combinational and optimized for a single pipeline stage. As such, these operations typically execute within a **single-cycle latency** in pipelined digital processors. The final scaling stage, which adjusts the Gaussian sample to match a target mean and variance, involves one additional 8-bit multiplier and one adder. This stage typically introduces an additional **1–2 ns** of delay.

Considering all functional stages, the **total latency** to generate one 8-bit Gaussian-distributed random number in 28nm CMOS is estimated to be approximately **8–15 ns**.

***Area efficiency:***

The pseudo-random number generator is most commonly realized using an 8-bit linear feedback shift register (LFSR) [9]. D flip-flops, XOR gates, and control logic require approximately 60 gate equivalents (GE), translating to around 90–120 $\mu m^2$ in TSMC 28nm standard-cell libraries. For the Box–Muller transformation, both the natural logarithm and square root functions can be implemented using small lookup tables (LUTs) or iterative logic, each requiring roughly 250–300 GE. Trigonometric operations such as sine and cosine are typically implemented using CORDIC algorithms [8], which require additional shift-and-add stages and control logic. These blocks can consume about 500–1000 $\mu m^2$ in total. Finally, to support scaling to arbitrary mean and standard deviation, the design includes an additional multiplier and an adder, contributing another 1000 $\mu m^2$.[10]

The total area required to implement a Gaussian random number generator in 28nm CMOS is approximately 0.0026 to 0.0032 $mm^2$.

| | Energy (pJ) | Time latency (ns) | Area($\mu m^2$) |
|---|---|---|---|
| Uniform RNG | 2 | 1-2 | 90-120 |
| Box-Muller | 115-225 | 7-11 | 800-2300 |
| Scaling | 3 | 1-2 | 1000 |
| Total | 119-229 | 8-15 | 2600-3200 |

**References**

[1] Spethmann, Jonas, et al. "Zero-field skyrmionic states and in-field edge-skyrmions induced by boundary tuning." Communications Physics 5.1 (2022): 19.

[2] Emori, S., Bauer, U., Ahn, SM. et al. Current-driven dynamics of chiral ferromagnetic domain walls. Nature Mater 12, 611–616 (2013).

[3] Pamula, V.R., Sun, X., Kim, S.M., ur Rahman, F., Zhang, B. and Sathe, V.S., 2019. A 65-nm CMOS 3.2-to-86 Mb/s 2.58 pJ/bit highly digital true-random-number generator with integrated de-correlation and bias correction. *IEEE Solid-State Circuits Letters*, *1*(12), pp.237-240.

[4] Wang, Y., Deng, D., Liu, L., Wei, S. and Yin, S., 2021, June. Lpe: Logarithm posit processing element for energy-efficient edge-device training. In *2021 IEEE 3rd International Conference on Artificial Intelligence Circuits and Systems (AICAS)* (pp. 1-4). IEEE.

[5] Wu, D., Chen, T., Chen, C., Ahia, O., San Miguel, J., Lipasti, M. and Kim, Y., 2019, July. SECO: A scalable accuracy approximate exponential function via cross-layer optimization. In *2019 IEEE/ACM International Symposium on Low Power Electronics and Design (ISLPED)* (pp. 1-6). IEEE.

[6] Aguirre-Hernandez, M. and Linares-Aranda, M., 2008, November. Energy-efficient high-speed CMOS pipelined multiplier. In *2008 5th International Conference on Electrical Engineering, Computing Science and Automatic Control* (pp. 460-464). IEEE.

[7] Osama, M., Gaber, L. and Hussein, A., 2016, February. Design of high performance Pseudorandom Clock Generator for compressive sampling applications. In *2016 33rd National Radio Science Conference (NRSC)* (pp. 257-265). IEEE.

[8] Andraka, R., 1998, March. A survey of CORDIC algorithms for FPGA based computers. In *Proceedings of the 1998 ACM/SIGDA sixth international symposium on Field programmable gate arrays* (pp. 191-200).

[9] Bagalkoti, A., Shirol, S.B., Kumar, P. and BS, R., 2019, February. Design and implementation of 8-bit LFSR, bit-swapping LFSR and weighted random test pattern generator: a performance improvement. In *2019 International Conference on Intelligent Sustainable Systems (ICISS)* (pp. 82-86). IEEE.

[10] Akhter, S., Saini, V. and Saini, J., 2017, February. Analysis of vedic multiplier using various adder topologies. In *2017 4th International Conference on Signal Processing and Integrated Networks (SPIN)* (pp. 173-176). IEEE.